\documentclass[sn-aps]{sn-jnl}

\usepackage{graphicx}%
\usepackage{subfig}
\usepackage{multirow}%
\usepackage{amsmath,amssymb,amsfonts}%
\usepackage{amsthm}%
\usepackage{mathrsfs}%
\usepackage[title]{appendix}%
\usepackage{xcolor}%
\usepackage{textcomp}%
\usepackage{manyfoot}%
\usepackage{booktabs}%
\usepackage{algorithm}%
\usepackage{algorithmicx}%
\usepackage{algpseudocode}%
\usepackage{listings}%

\begin{document}

\title[Article Title]{New Chinese Facilities for Short-Range Correlation Physics}


\author*{\fnm{Zhihong} \sur{Ye}}\email{yez@tsinghua.edu.cn}
\author{\fnm{Haojie} \sur{Zhang}}
\author{\fnm{Yaopeng} \sur{Zhang}}
\author{\fnm{Haocen} \sur{Zhao}}

\affil{\orgdiv{Department of Physics}, \orgname{Tsinghua University}, \orgaddress{\city{Beijing}, \postcode{100084},\country{China}}}


\abstract{This article explores the significant advancements in Short-Range Correlation (SRC) research enabled by the latest Chinese nuclear physics facilities—CSR at HIRFL, HIAF, SHINE, and the upcoming EicC. These facilities introduce cutting-edge technologies and methodologies, addressing existing challenges and broadening the scope for SRC studies. By providing detailed insights into the capabilities and expected contributions of each facility, the paper highlights China's emerging role in the global nuclear physics landscape. The collaborative potential, alongside complementary global efforts, positions these facilities to deeply influence our understanding of nuclear matter's fundamental properties and interactions.}


\maketitle

\section{Introduction}\label{sec1}

Short-Range Correlations (SRC) play a critical role in understanding nucleon interactions within the compact confines of atomic nuclei~\cite{CiofidegliAtti:2015lcu,Hen:2016kwk}. The initial detection of SRC at the Stanford Linear Accelerator Center (SLAC)~\cite{Frankfurt:1988nt,Frankfurt:1993sp} and Brookhaven National Lab (BNL)~\cite{Tang:2002ww,Piasetzky:2006ai} marked a significant milestone in nuclear physics research. Follow-up high-precision quasi-elastic electron scattering (QES) experiments at the Thomas Jefferson Laboratory (JLab) have deepened our understanding by highlighting the dominance of neutron-proton pairs in two-nucleon SRC (2N-SRC) states. These findings have not only enhanced our grasp of nucleon-nucleon (NN) interactions but also set new standards for theoretical models in the field~\cite{Arrington:2011xs,Fomin:2017ydn, Arrington:2022sov}.

Experiments on fixed nucleus targets have encountered limitations such as kinematic distortions and strong final state interactions (FSI), which impact nucleon reconstruction and the detection of the A-2 fragment post-SRC pair ejection. Overcoming FSI and accurately probing ultra-high-momentum nucleons from 3-nucleon SRC (3N-SRC) clusters necessitates significantly higher energy transfers. This requirement constrains the valid kinematic region and reduces QES cross-sections. While 2N-SRC has been extensively studied at JLab, evidence for 3N-SRC remains elusive~\cite{Ye:2017ivm}. Additionally, the limited availability of stable asymmetric nuclear targets restricts systematic exploration of the isospin effect in SRC pair formation, crucial for understanding NN interactions~\cite{Subedi:2008zz,Hen:2014nza, CLAS:2019vsb,CLAS:2020mom,Li:2022fhh}.

\begin{figure}[ht!]
\begin{center}
\includegraphics[width=0.8\textwidth]{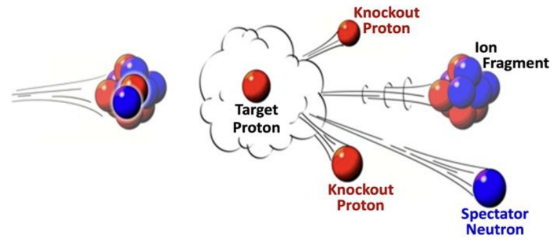}
\caption{Illustration of the proton-nucleus collision in the inverse kinematic}
    \label{fig:invserpA}
\end{center}
\end{figure}

To address the limitations of fixed-target experiments, facilities like BM@N at JINR and R3B at GSI have embraced inverse kinematics, where an ion beam collides with a stationary hydrogen target (called inverse-pA in short), see Fig.~\ref{fig:invserpA}, significantly reducing FSI, improving the detection of high-momentum nucleons and enabling the clean separation of the A-2 fragment from beam ions and other decayed fragments. This method also greatly increases reaction cross-sections due to the strong interaction in proton-nucleus collisions, offering a stark contrast to electron-nucleus scattering dominated by electromagnetic interactions. Despite the innovative approach of BM@N using 10 GeV/u $^{12}$C ion beams for SRC research~\cite{BMN:2021yag}, challenges persist in ion species diversity, beam intensity, and detection efficiency. R3B at GSI has contributed insights into SRC in asymmetric nuclei, like $^{16}$C, challenging to probe in fixed-target experiments due to lower ion beam energy (~1.0 GeV/u), which leads to data contamination from mean-field nucleon knockouts or SRC events with significant FSI effects, further compounded by beam quality and detector coverage limitations.

The introduction of cutting-edge Chinese facilities marks a transformative era for SRC research, enhancing precision and overcoming existing infrastructure limitations. These facilities, armed with advanced technology, significantly improve our ability to delve into nucleon interactions with unparalleled detail. Complementing these advancements, China's theoretical and experimental SRC communities have seen rapid growth. The first workshop on SRC research in China, held in Huizhou, Guangdong in November 2023, brought together more than 80 scientists from around the world, fostering discussions that underpin the subsequent sections of this article~\cite{src-china-2023}.

\section{CEE Program at HIRFL-CSR}\label{sec-cee}
\begin{figure}[ht!]
\begin{center}
\includegraphics[width=0.8\textwidth]{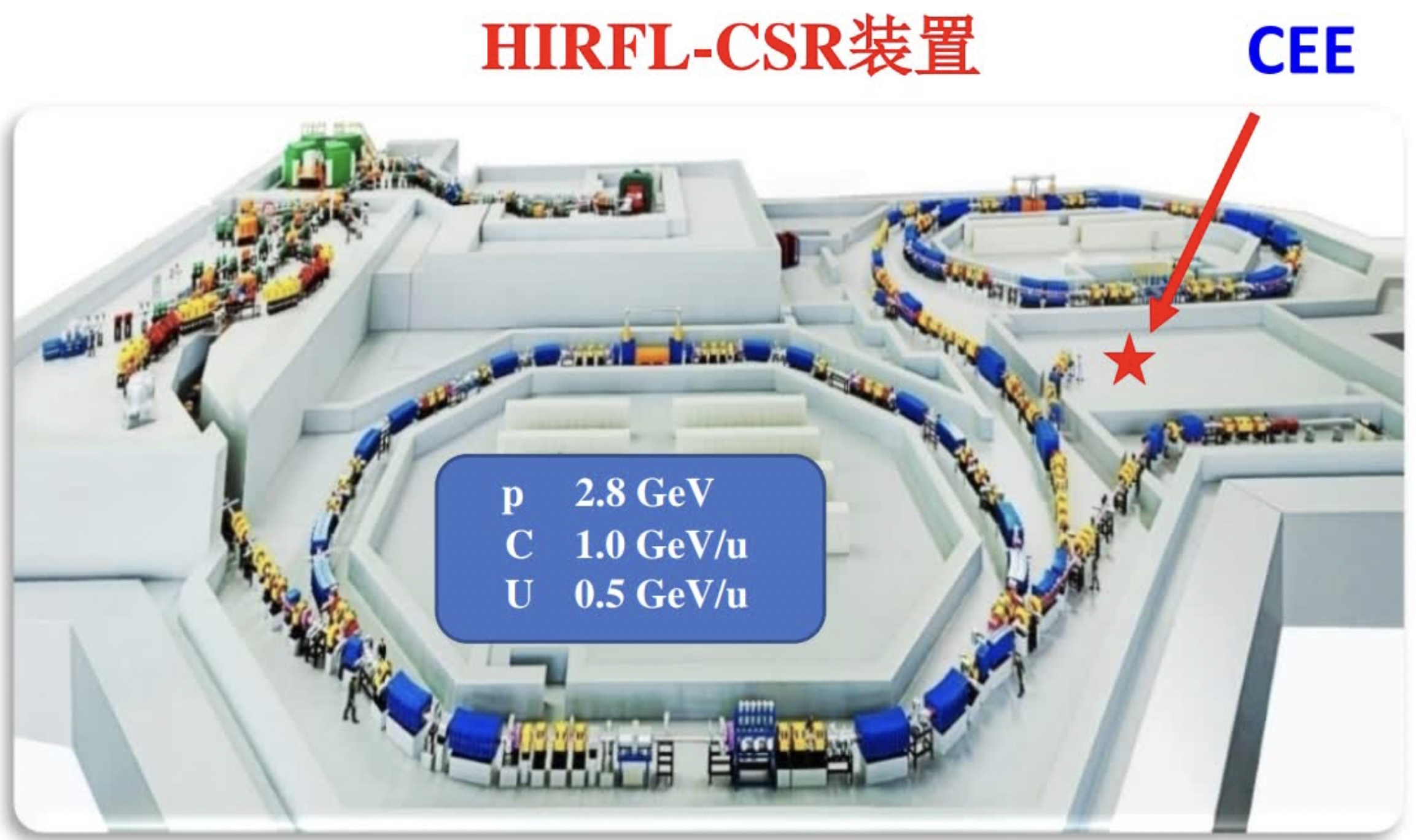}
\caption{HIRFL-SRC in IMP and the location of the CEE detector}
    \label{fig:csr}
\end{center}
\end{figure}
Located within the Institute of Modern Physics (IMP) in Lanzhou, Gansu, China, the Heavy Ion Research Facility in Lanzhou (HIRFL) exemplifies cutting-edge nuclear physics research~\cite{Mao2020IntroductionOT}. Its Cooling Storage Ring (CSR), shown in Fig.~\ref{fig:csr}, stands out for its remarkable flexibility and capacity, handling a broad spectrum of ion beams from light hydrogen to heavy uranium ions. This enables a wide range of experiments, with beam energies up to 2.8 GeV for protons and 1.0 GeV per nucleon (GeV/u) for light ions, extending to several hundred MeV/u for the heaviest ions, making CSR a pivotal facility for high-precision investigations of nuclear structure~\cite{Xia2002TheHI,Yuan2013StatusOT}.

\begin{figure}[ht!]
\begin{center}
\includegraphics[width=0.7\textwidth]{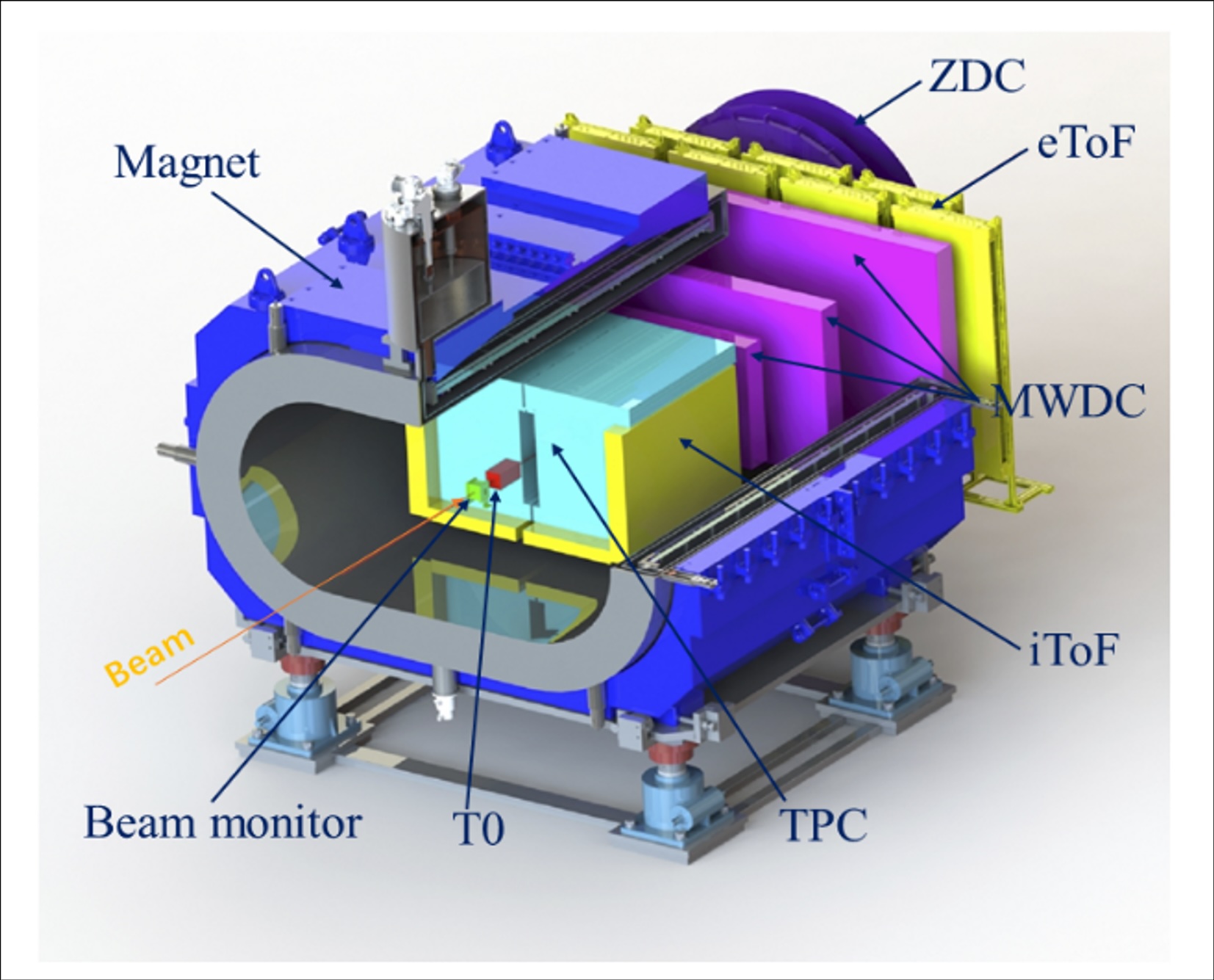}
\caption{Conceptual design of the CEE detector system}
    \label{fig:cee}
\end{center}
\end{figure}

At the core of HIRFL-CSR's innovation is the CSR External-Target Experiment (CEE) detector, a tool equipped for high-resolution and comprehensive coverage, essential for exploring intricate nuclear phenomena~\cite{Xiao2013ProbingNS, Lu:2016htm}. It ambitiously aims to probe key nuclear physics questions, such as the quantum chromodynamics (QCD) phase diagram's critical point at low temperatures and high densities, the nuclear matter equation of state, hypernucleus structure, and nucleon-nucleon interactions. The CEE detector stands out for its ability to manage event rates of 10 kHz with almost complete angular coverage, excelling in particle identification for charged particles (e.g. $\pi^{\pm}$, $K^{\pm}$, $p$, $d$, $t$, $^3$He and $^4$He) with a momenta measurement precision of about 5\%.

As shown in Fig.~\ref{fig:cee}, this detector system is notable for its large-scale magnet, dimensioned at 3.4m (L) x 3.2m (W) x 1.6m (H) with a field strength of 0.5T, optimized for heavy-ion collision detection. The Time-Projection Chamber (TPC) housed within the magnet is engineered for precise particle position (500$\mu$m resolution) and momentum (5\% accuracy) measurements, alongside efficient two-track separation down to 3cm. Surrounding the TPC are three inner time-of-flight (iTOF) planes aiming for a timing resolution of 50ps and detection efficiencies above 95\%. Additionally, three layers of multiple-wire drift chambers (MWDC) offer charged particle tracking with resolutions up to 300$\mu$m and energy resolution of 22\%, maintaining detection efficiencies above 98\%. An external TOF detector (eTOF) positioned behind the magnet and a zero-degree calorimeter (ZDC) further enhance the system's measurement capabilities, with the ZDC focusing on particle charge determination with a 15\% resolution for charges up to 15. The construction of the CEE detector system is progressing as planned, with assembly set for completion by the end of 2024, followed by commissioning and data collection in 2025.

\begin{figure}[ht!]
\begin{center}
\includegraphics[width=0.8\textwidth]{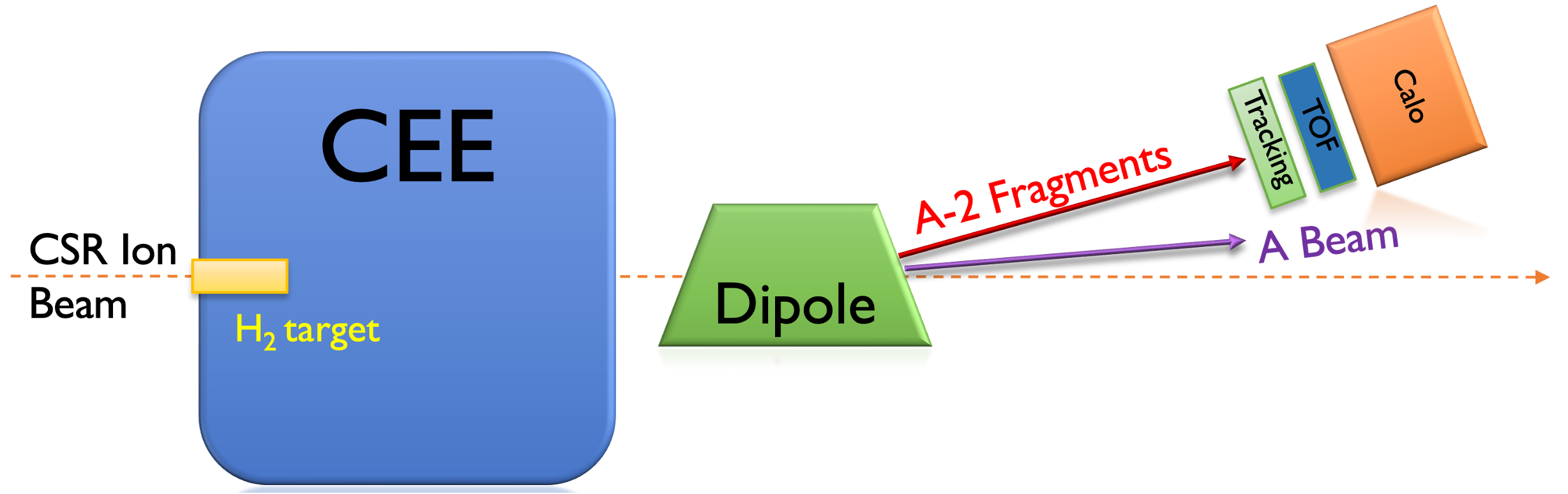}
\caption{Sketch of CEE with additional fragment detectors for SRC measurement}
    \label{fig:cee-src}
\end{center}
\end{figure}

\begin{figure}[ht!]
\begin{center}
\includegraphics[width=0.9\textwidth]{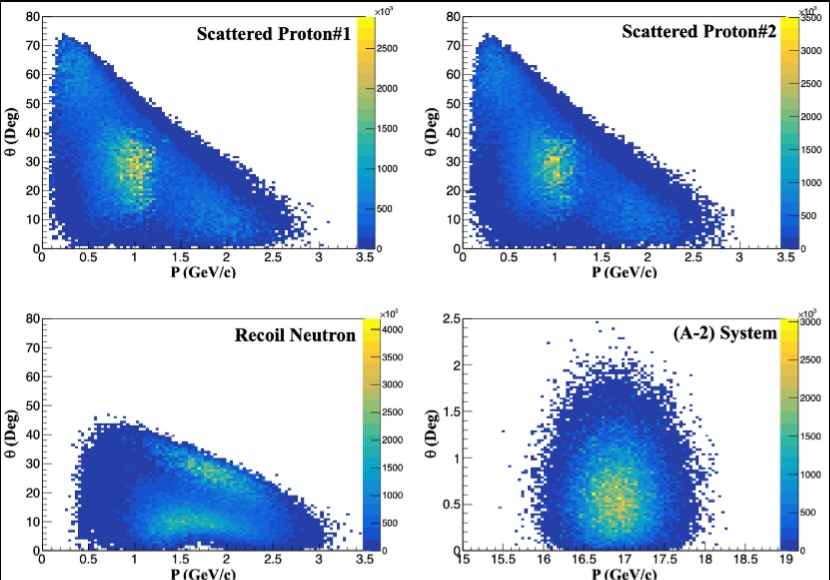}
\caption{Distribution of scattered angles and momenta of all final state particles in inverse-pA SRC reaction using 1.0GeV/u $^{12}$C beam}
    \label{fig:cee-src-sim}
\end{center}
\end{figure}

The HIRFL-CSR's ion beam energy, closely aligned with GSI's at approximately 1.0 GeV/u for $^{12}$C, facilitates precise SRC measurements via inverse pA reactions, focusing on nucleon momentum distribution transitions from mean-field to SRC tail. The default nuclear targets will be replaced by a liquid hydrogen target for SRC research adaptation, enhancing experimental accuracy. Monte-Carlo simulations, based on GCF theory~\cite{Weiss:2015mba}, illustrate the neutrons, protons, and A-2 system's angular and momentum ranges post-collision (see Fig.~\ref{fig:cee-src-sim}), indicating the scattered protons fall within the CEE detector's acceptance. Before the neutron detection capability to be added to CEE, an additional fragment detection system is deemed necessary for precise SRC event isolation. This proposed system, as illustrated in Fig.~\ref{fig:cee-src}, potentially including a second dipole magnet, TOF planes, trackers, and a calorimeter, is under optimization through ongoing Geant4 simulations, showcasing the project's commitment to refining SRC measurement strategies with the CEE and its auxiliary detectors.

\section{High Intensity heavy ion Accelerator Facility - HIAF}\label{sec-hiaf}

\begin{figure}[ht!]
 \begin{center}
 \subfloat[Design sketch of HIAF]{
\includegraphics[width=0.8\textwidth]{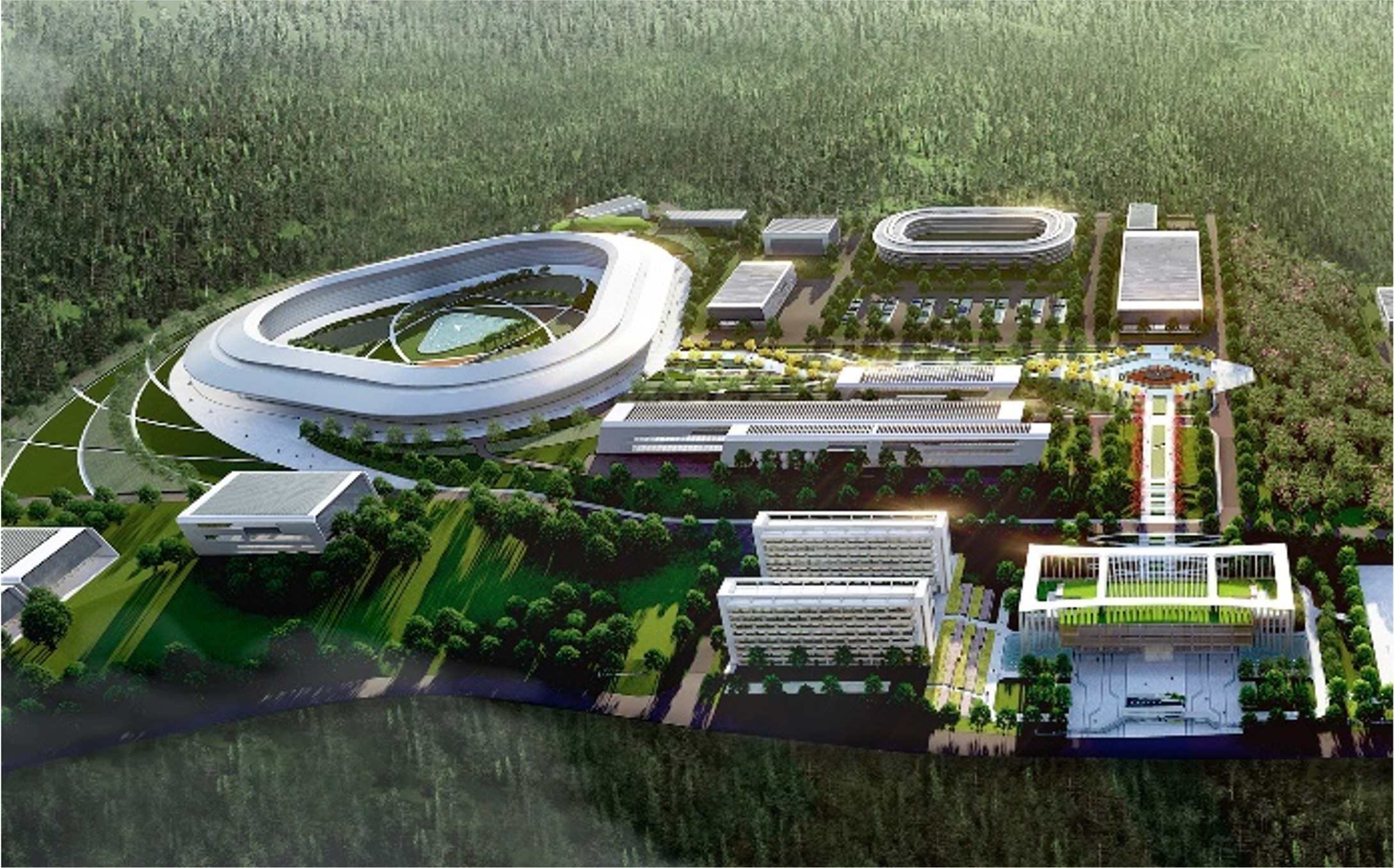} } \\
 \subfloat[The satellite image of HIAF's construction site as November 2023]{
  \includegraphics[width=0.8\textwidth]{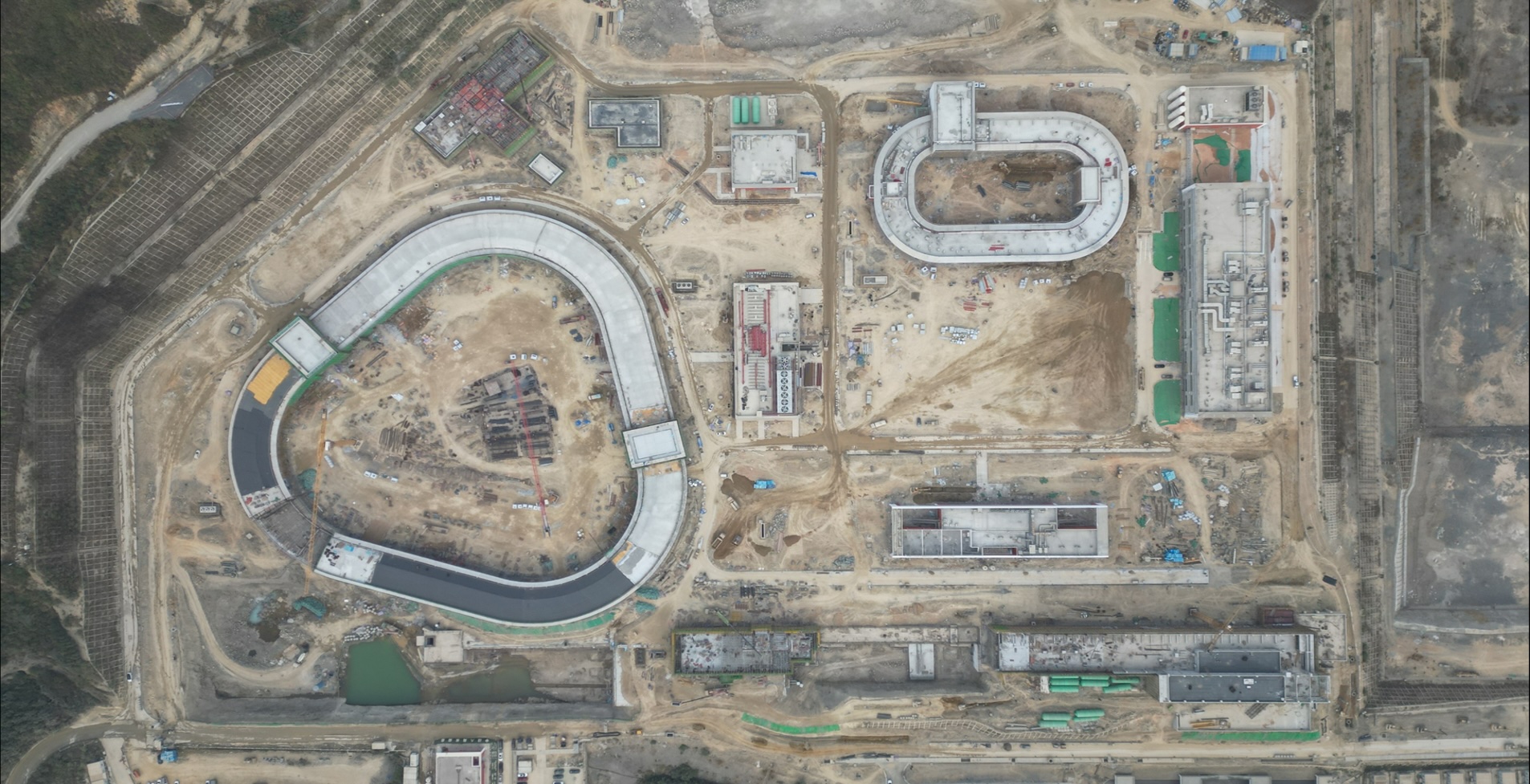} } 
     \caption{The High Intensity heavy ion Accelerator Facility (HIAF)}
    \label{fig:hiaf-design}
  \end{center}
\end{figure}

The High Intensity heavy ion Accelerator Facility (HIAF), under construction in Huizhou, Guangdong, and hosted by IMP, marks a significant leap in nuclear physics and technology~\cite{Yang2013HighIH, Yang:2023hfx,Ma2017HIAFNO}. The layout and current construction progress of HIAF are shown in Fig.~\ref{fig:hiaf-design}. HIAF is set to address crucial questions in nuclear physics, including the origin of elements, few-body nucleon-nucleon interactions, exotic nuclear structures, the limits of nuclear existence, and the properties of high-energy density nuclear matter.

\begin{figure}[ht!]
\begin{center}
\includegraphics[width=0.95\textwidth]{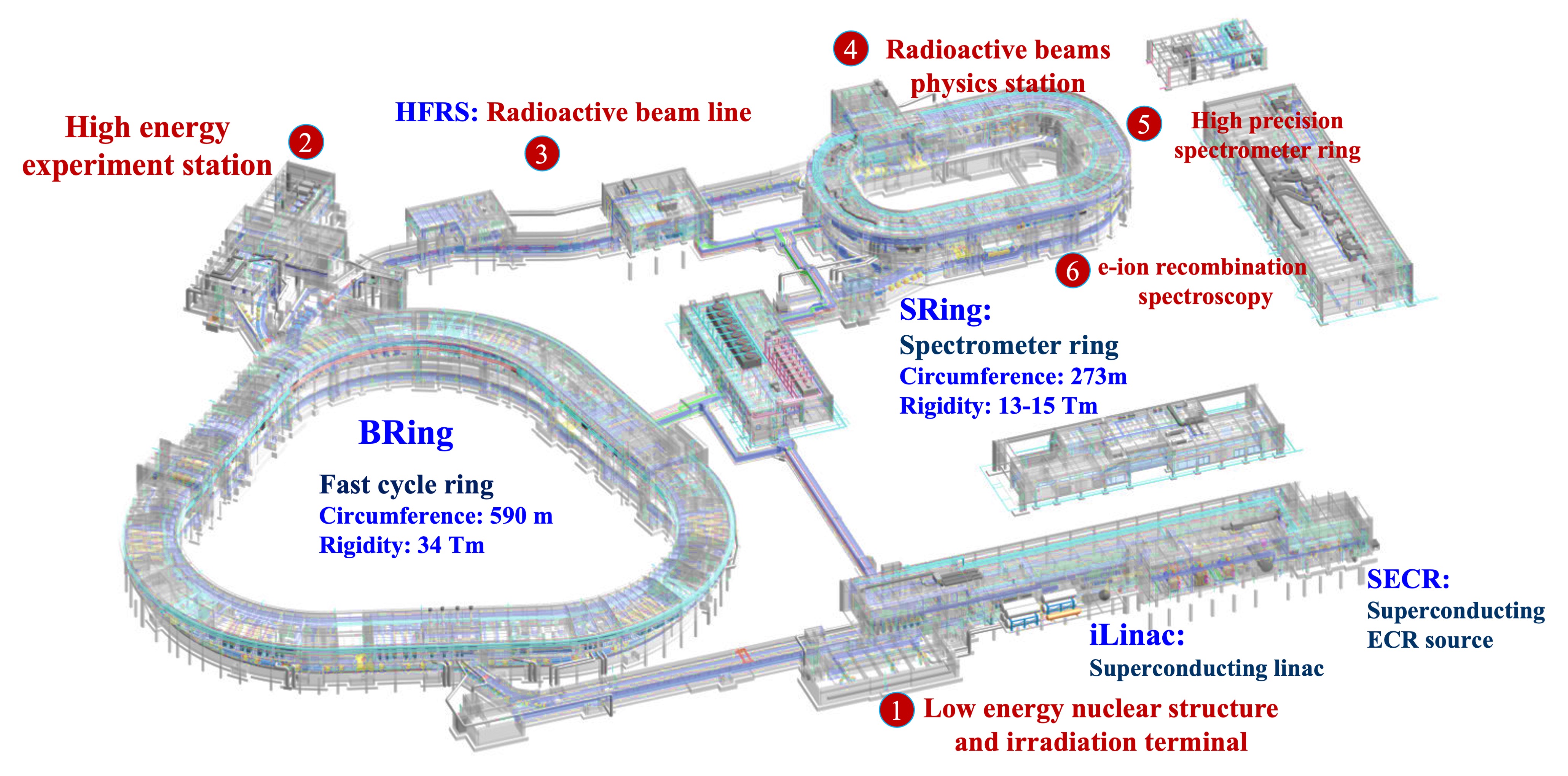} 
\caption{The layout of accelerator complexes of HIAF}
    \label{fig:hiaf-design2}
\end{center}
\end{figure}

Shown in Fig.~\ref{fig:hiaf-design2}, HIAF is equipped with a high-power linear accelerator, a booster ring, and a main storage ring, and it is engineered for exceptional heavy-ion acceleration performance. The linear accelerator, notable for its ion beam delivery of up to 100MeV/u for heavy ions with an intensity of 28p$\mu$A, employs advanced RF technology and beam focusing mechanisms for unparalleled precision. With a magnetic rigidity of up to 34Tm, the booster ring complements this capability by facilitating ion acceleration up to 10GeV/u for ions as heavy as U$^{35+}$ at maximum intensity of 2$\times$10$^{11}$ particles per pulse (ppp). The HFRS (FRagment Separator of HIAF) is designed to segregate various secondary fragments, primarily short-lived radioactive nuclei, facilitating their collision with a fixed heavy-element target and linking the BRing to the SRing with a magnetic rigidity of up to 25Tm. At the heart of HIAF, the main storage ring features a maximum magnetic rigidity of 15Tm and is outfitted with advanced cooling systems and diagnostic tools, ensuring conditions conducive to groundbreaking research with beam intensities reaching up to 1$\times$10$^{12}$ ppp.

HIAF will host a suite of experimental stations to support a broad spectrum of physics programs. Among these, the High-Energy-Station (HES) is designated for high-energy nuclear physics experiments, utilizing proton beams of up to 9.3GeV/u (or 2.45GeV/u for U$^{35+}$ beams) with intensity of 1$\times$10$^{11}$ extracted from the BRing. Ion beams directed to HES will be split into three branches, each serving different fixed-target experimental systems, including the relocation and upgrade of the CEE detector from HIRFL-CSR.

\begin{figure}[ht!]
 \begin{center}
    \includegraphics[width=0.7\textwidth]{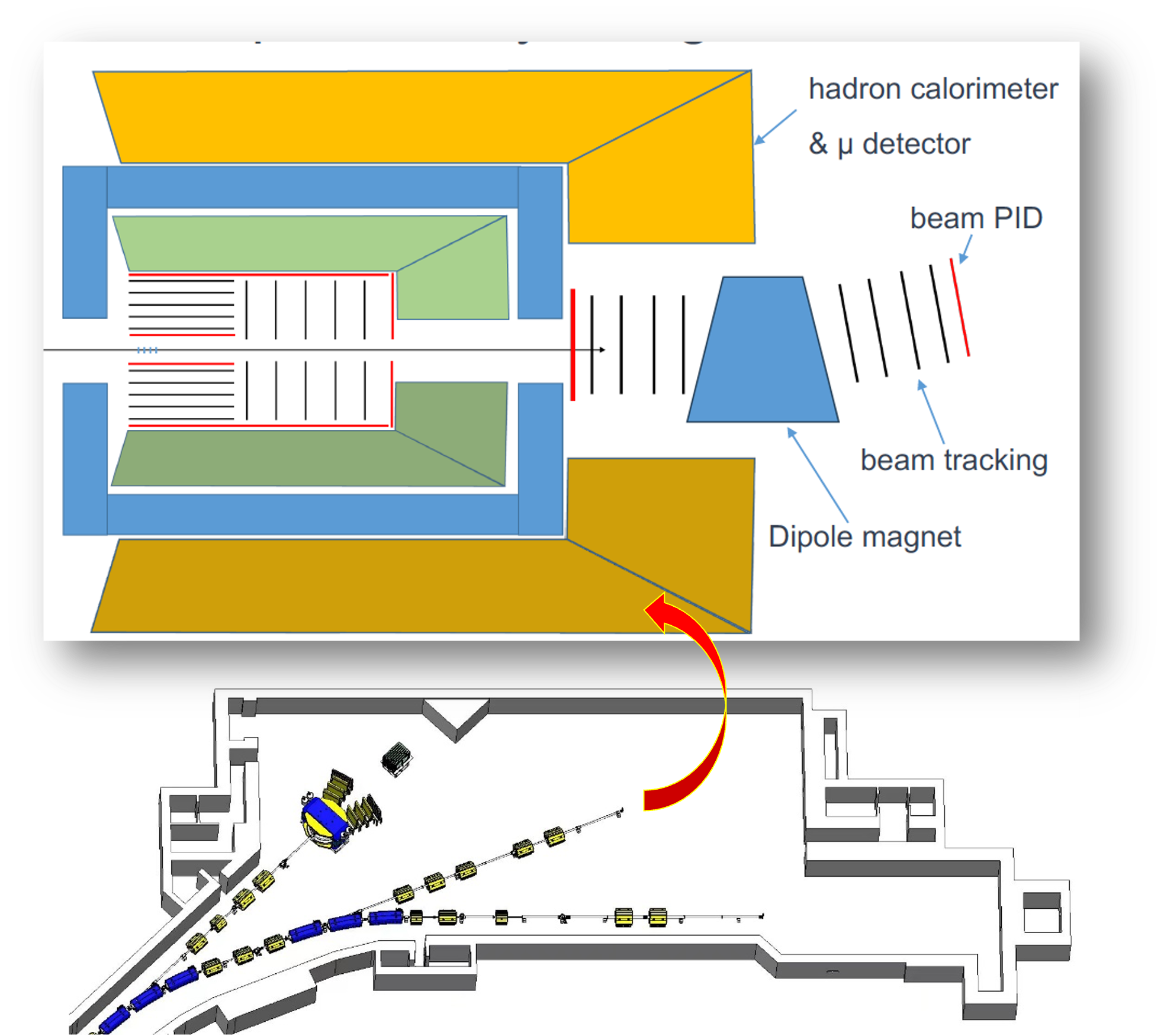} 
   \caption{Sketch of the detector system in the high-energy station}
    \label{fig:hiaf-src}
  \end{center}
\end{figure}

\begin{figure}[ht!]
\begin{center}
\includegraphics[width=0.9\textwidth]{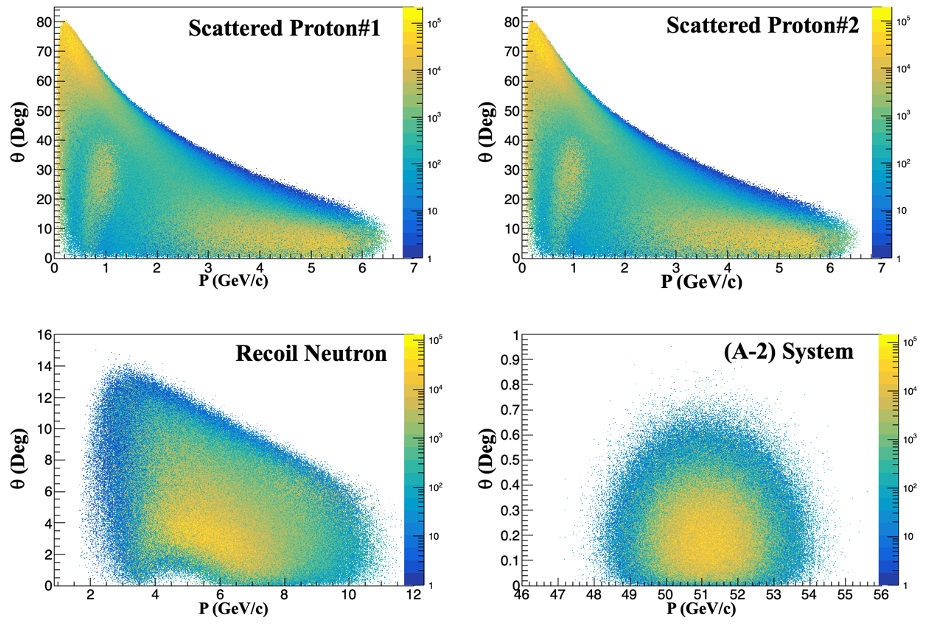}
\caption{Distribution of scattered angles and momenta of all final state particles in inverse-pA SRC reaction using 9.3GeV/u $^{12}$C beam from HIAF}
    \label{fig:hiaf-src-sim}
\end{center}
\end{figure}

A brand new general-purpose detector system planned for HES features a 1.0T solenoid magnet, silicon-based trackers, TOFs, and calorimeters to detect high-momentum particles ($\pi^{\pm}$ ,  $K^{\pm}$, protons and light ions) at a high rate (see Fig.~\ref{fig:hiaf-src}). Surrounding the magnet is the muon and neutron detection system. For the inverse-pA SRC measurement, the scattered protons and neutrons will be measured by this detector system. In addition, a downstream small-angle detector system will identify A-2 fragments, enhancing the setup with a dipole, tracking layers, TOF planes, and calorimeters. This comprehensive full acceptance detector system, with exceptional beam quality, is set to enable high-precision SRC measurements, offering extensive statistics and broad kinematic coverage vital for detailed 2N-SRC studies and exploring elusive 3N-SRC signals.  Assuming 100\% acceptance in this detector system, Fig.~\ref{fig:hiaf-src-sim} shows the angular and momentum ranges of all final state particles from the SRC reactions with 9.3 GeV/u $^{12}$C beam. Ongoing simulations are fine-tuning the detector's design for optimal performance.

HIAF dedicates several experimental stations along the HFRS for investigating nuclear reaction mechanisms and the structure of rare radioisotopes, with a special emphasis on studying SRC pair formation in large asymmetric nuclei. This focus addresses a gap in QES experiments, which are limited by the absence of stable, large-asymmetric nuclear targets for such studies.

Scheduled for completion by the end of 2024 with beam commissioning in 2025, HIAF is set to begin its first experiments by the year's end. The HIAF upgrade (HIAF-U) project plans to increase proton beam energies to 20 GeV/u (or 9.1 GeV/u for U$^{35+}$), aiming to improve beam intensity and quality through superconducting accelerator technology.

The inauguration of HIAF marks a significant advancement in global nuclear science collaboration, promising to enhance both fundamental and applied research worldwide. With several physics programs in development and increasing opportunities for collaboration, HIAF is set to become a key contributor to SRC physics, propelling experimental research into new realms of precision.

\section{Electron-Ion Collider in China - EicC}\label{sec-eicc}
\begin{figure}[ht!]
\begin{center}
\includegraphics[width=0.95\textwidth]{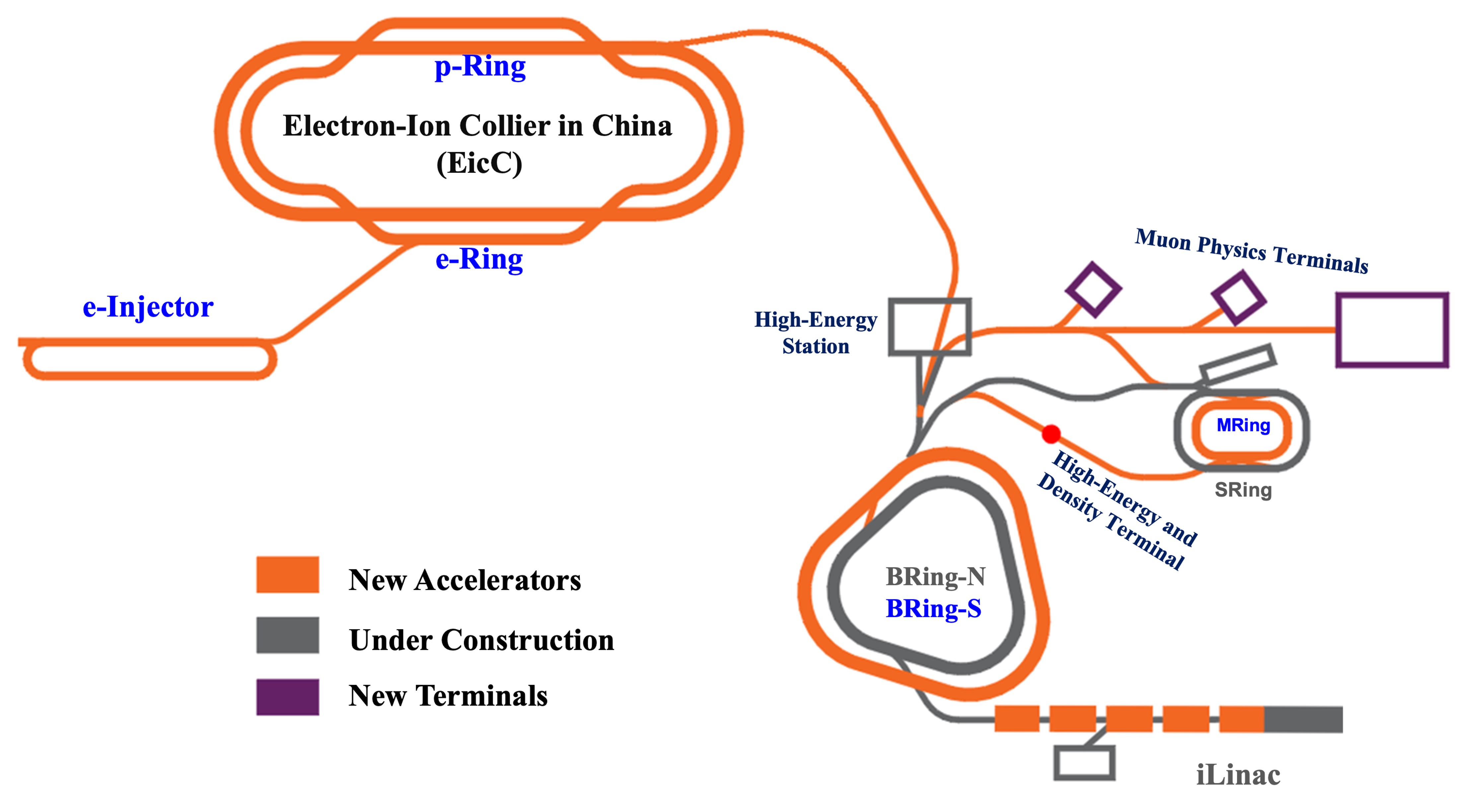}
\caption{The preliminary design of the Electron-Ion Collider in China (EicC)}
    \label{fig:eicc}
\end{center}
\end{figure}

\begin{figure}[ht!]
\begin{center}
\includegraphics[width=0.95\textwidth]{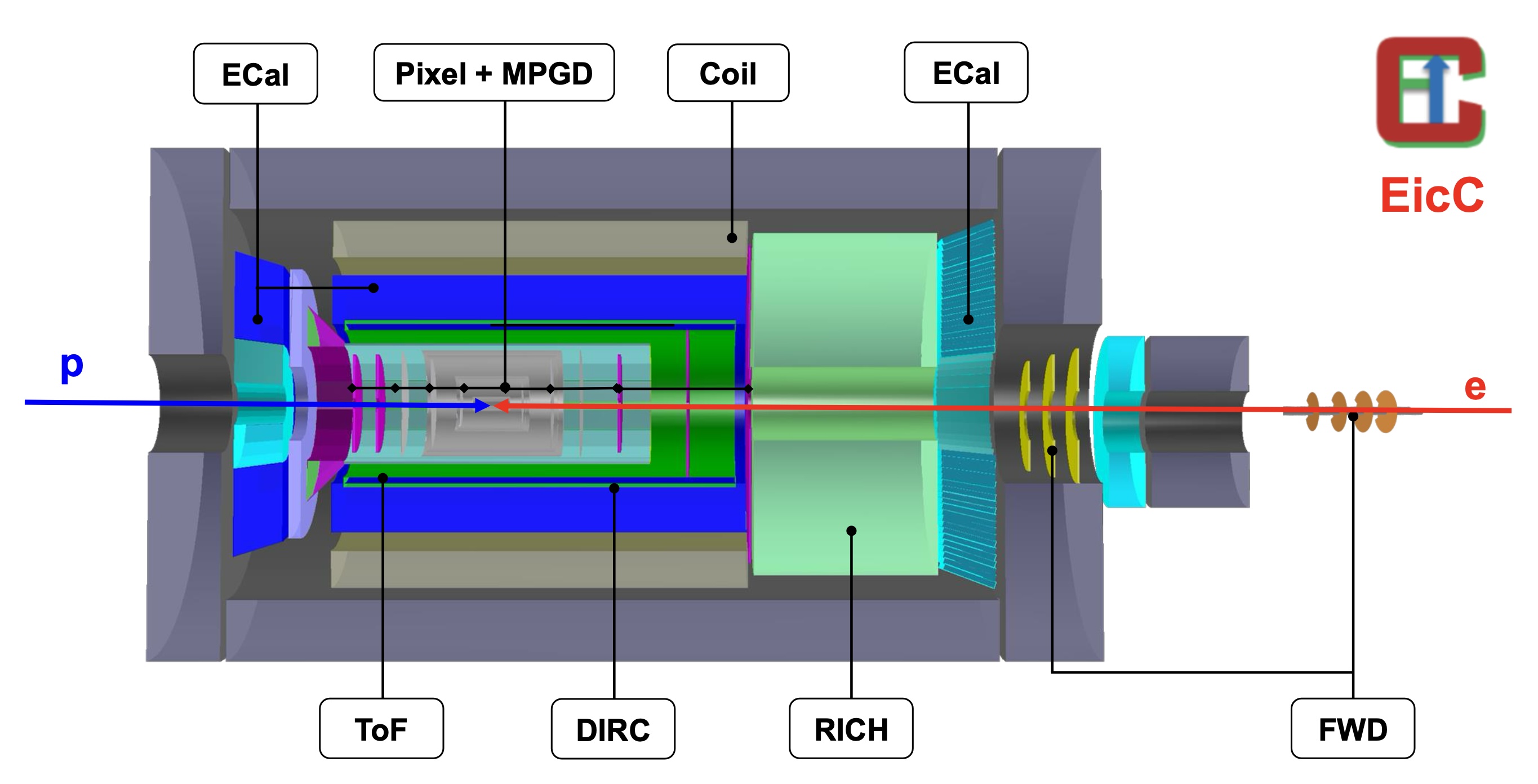}
\caption{The preliminary design of the first detector system on EicC}
    \label{fig:eiccdet}
\end{center}
\end{figure}

The forthcoming upgrade of HIAF heralds the development of the Electron-Ion Collider in China (EicC), introducing a state-of-the-art electron accelerator and a collider ring into the existing framework~\cite{Anderle:2021wcy}. Fig.~\ref{fig:eicc} reveals the layout of the future EicC. This upgrade envisions a comprehensive accelerator complex, featuring an injector ring capable of delivering up to 3.5 GeV polarized electron beams, alongside two dedicated collider rings—one for electrons (eRing) and another for protons and ions (pRing). In alignment with the HIAF-U enhancements, the ion injector will be upgraded to produce polarized proton, $^2$D, and $^3$He ion beams, with the BRing supplying 20GeV/u proton or ion beams to the pRing through the HES. The layout includes four interaction points (IPs), each accommodating different detector systems, with one currently in the design phase (see Fig.~\ref{fig:eiccdet}). A conceptual design report of EicC is under preparation and will be released by the end of 2024.

EicC aims to spearhead research into the QCD of nuclear matter, particularly focusing on the exploration of sea quark components within nucleons and nuclei. This initiative complements the efforts of the future US-EIC at BNL~\cite{AbdulKhalek:2021gbh}, filling a crucial gap in the kinematic range between JLab and the US-EIC. EicC's unique capabilities will enable detailed investigations into SRC physics through tagged quasi-elastic and deep-inelastic scattering processes~\cite{Hauenstein:2021zql}, as well as diffractive J/$\psi$ production~\cite{Tu:2020ymk} . Furthermore, the facility will facilitate systematic studies on the interplay between SRC and modifications of the parton structure in nuclei~\cite{Weinstein:2010rt}, addressing puzzling phenomena such as the EMC effect and anti-shadowing~\cite{Gomez:1993ri, Geesaman:1995yd}, thereby enriching our understanding of nuclear matter's underlying QCD dynamics.

\section{Opportunities at SHINE}\label{sec-shine}

\begin{figure}[ht!]
\begin{center}
\includegraphics[width=0.95\textwidth]{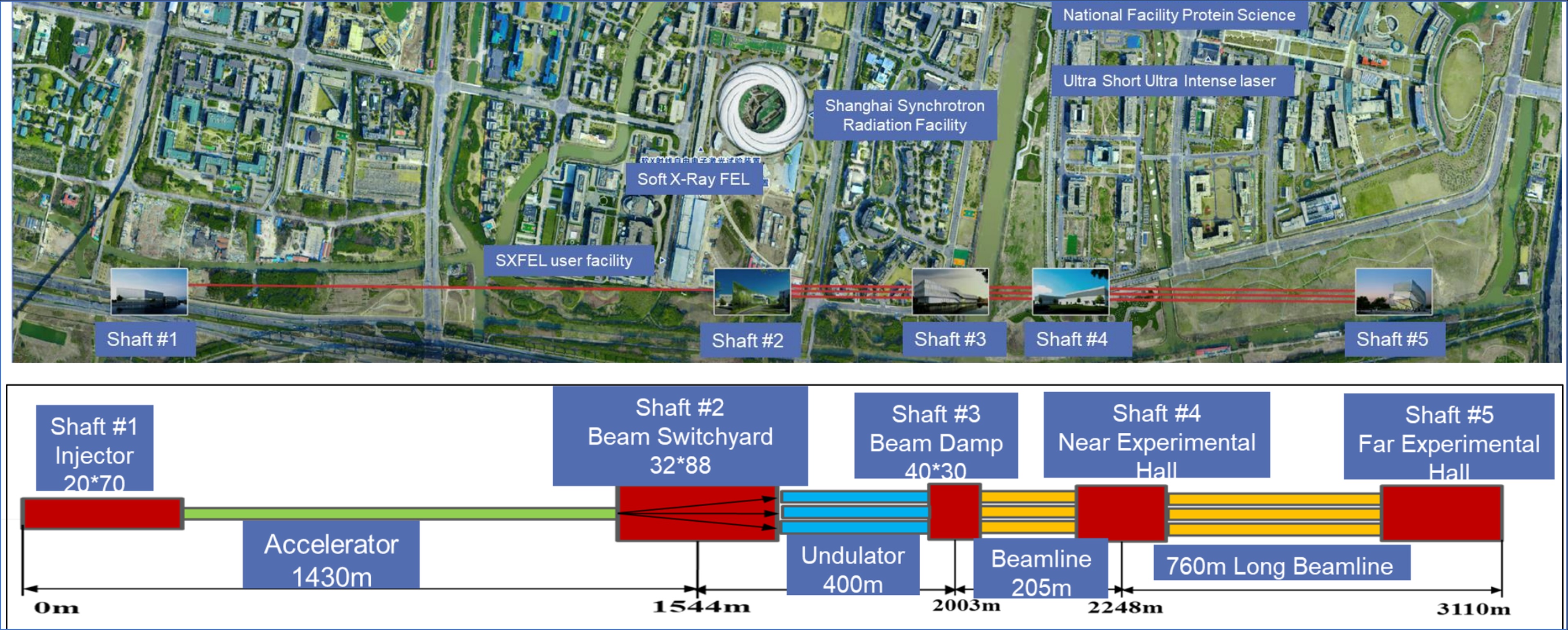}
\caption{Satellite image of SHINE's surrounding and its layout of beamlines and shafts}
    \label{fig:shine}
\end{center}
\end{figure}

The Shanghai High Repetition Rate X-ray Free Electron Laser and Extreme Light Facility (SHINE) marks a significant advancement in nuclear science, offering unparalleled capabilities for generating intense X-ray and optical light beams~\cite{Zhao:2017ood,Zhao:2018lcl}. At its core, SHINE's Free Electron Lasers (FELs) provide a wide range of electromagnetic radiation, with X-rays of exceptional resolution for atomic and molecular studies. Its high repetition rate enhances SHINE's capacity for swift, high-quality measurements, making it an invaluable asset across scientific disciplines, from fundamental particle physics to materials science.

Positioned alongside the prestigious Shanghai Synchrotron Radiation Facility and the newly established soft X-Ray FEL, as shown in Fig.~\ref{fig:shine}, SHINE epitomizes the forefront of scientific infrastructure. Its mainstay, a 1.43km long linear superconducting electron accelerator, resides in a 29m deep underground tunnel. This accelerator generates high-intensity electron beams with energies up to 8.6GeV, bunch charges from 10pC to 300pC, beam power as high as 2.5MW, pulse widths ranging from 5fs to 200fs, and operates at a radio frequency of 1MHz~\cite{Cao:2020wad}.

\begin{figure}[ht!]
\begin{center}
\includegraphics[width=0.95\textwidth]{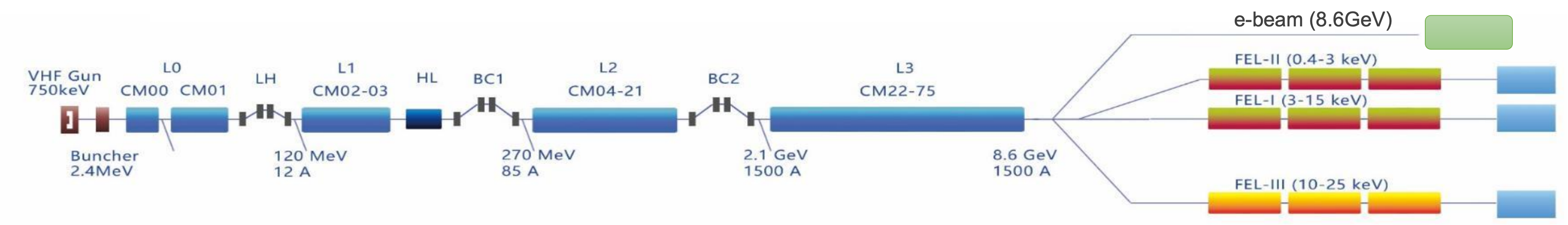}
\caption{SHINE's accelerator components, its electron line and three FEL lines}
    \label{fig:shine2}
\end{center}
\end{figure}

Accessible through five shafts housing essential beam devices, the beam switchyard in the second shaft channels the electron beam into four branches. Three undergo a series of undulators to generate FELs across different energy ranges, while the fourth maintains 8.6GeV electrons for various applications, as shown in Fig.~\ref{fig:shine2}. The last two shafts are designated for ten experimental stations, broadening SHINE's experimental scope.

\begin{figure}[ht!]
\begin{center}
\includegraphics[width=0.95\textwidth]{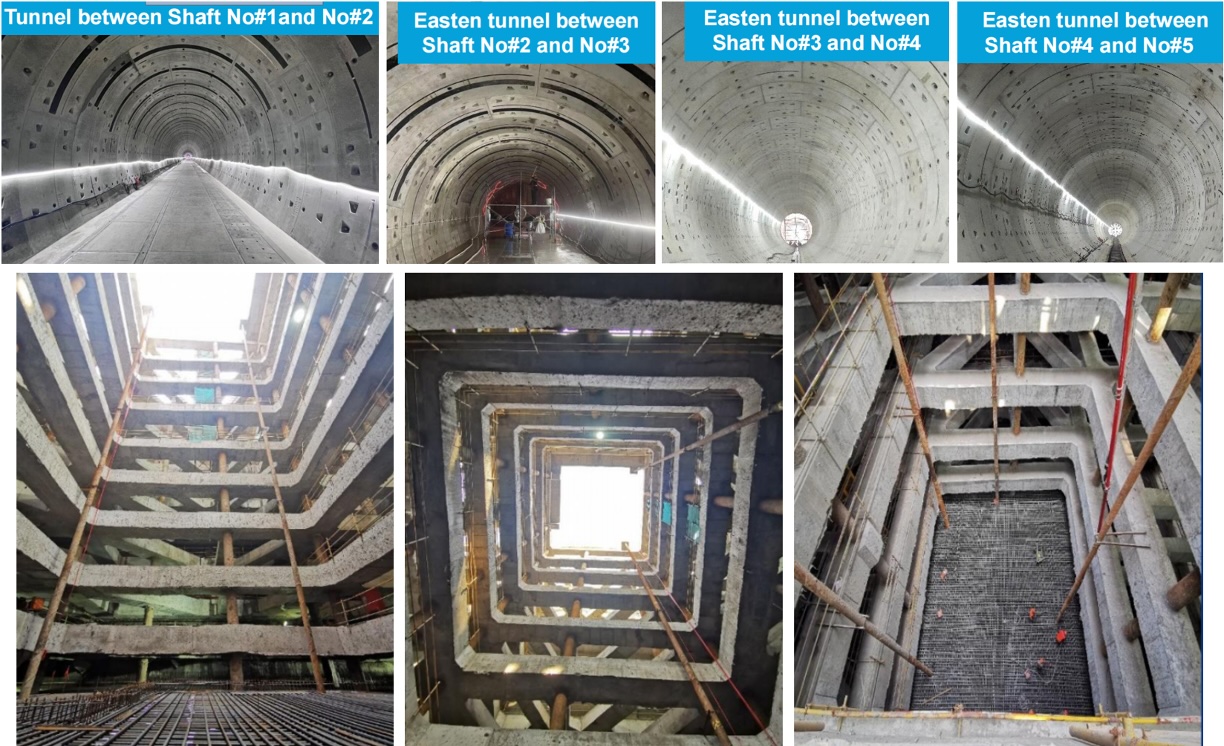}
\caption{Construction sites of SHINE's tunnels and three shafts as March 2023}
    \label{fig:shine3}
\end{center}
\end{figure}

As the civil construction approaches completion(see Fig.~\ref{fig:shine3}), SHINE is swiftly advancing towards installing its beam components, with the first FEL expected to be operational by the end of 2025.

SHINE's high-precision electron beams are set to explore quark dynamics within nucleons and their interactions within nuclei, akin to research in Hall-A, Hall-B CLAS12, and Hall-C at JLab. Moreover, SHINE's potent photon beams may facilitate photo-nuclear reaction studies, providing a new perspective on nucleon correlations and nuclear structure. This complements SRC research analogous to the GlueX project in Hall-D at JLab~\cite{GlueX:2020dvv}. Ongoing discussions within the Chinese nuclear physics community aim to develop experimental programs at SHINE, underscoring its potential to make substantial contributions to the SRC physics.

\section{Conclusion}\label{sec-conclusion-final}
The advancements in SRC esearch with new Chinese facilities - CSR at HIRFL, HIAF, SHINE, and the upcoming EicC - represent a significant leap in nuclear physics. These developments are set to complement existing physics programs globally, addressing current research challenges and enhancing the understanding of nuclear matter. By leveraging state-of-the-art technologies and promoting international collaboration, these facilities are poised to lead innovative investigations in SRC and beyond, marking a pivotal moment for the nuclear physics community and paving the way for groundbreaking discoveries in the fundamental forces within nuclei.

\backmatter
\bmhead{Acknowledgments}
The author gratefully acknowledges Gongtao Fan, Chang-bo Fu, Julian Kahlbow, Hongna Liu, Elizer Piasetzky, Maria Patsyuk, Hao Qiu, Zhigang Xiao, Jian-cheng Yang, Ya-Peng Zhang, and many colleagues for their valuable discussions and support. Special thanks to the National Natural Science Foundation of China under Grant Nos. 12275148 and 12361141822 and by Tsinghua University Initiative Scientific Research Program, and collaborators from various institutions for their essential roles in developing and operating the facilities discussed. The author also acknowledges the assistance of ChatGPT4.0 in preparing the article.

\bibliography{sn-bibliography}

\end{document}